\documentstyle[12pt]{article}
\def\eqs{Eqs. \hspace{-1.2 mm}}
\def\eq{Eq. \hspace{-1.2 mm}}
\begin{document}

\title{Self-Similar Collapse of Conformally Coupled Scalar Fields}
\author{H. P. de Oliveira and E. S. Cheb-Terrab \\
\     \\
{\em Departamento de  F\'{\i}sica Te\'orica}  \\
{\em Instituto de F\'{\i}sica}  \\
{\em Universidade do Estado do Rio de Janeiro}  \\
{\em RJ 20550 - Brazil}}
\date{}
\maketitle
\thispagestyle{empty}

\begin{abstract}

A massless scalar field minimally coupled to the gravitational field in a
simplified spherical symmetry is discussed. It is shown that, in this
case, the solution found by Roberts\cite{roberts}, describing a scalar
field collapse, is in fact the most general one. Taking that solution as
departure point, a study of the gravitational collapse for the
self-similar conformal case is presented.

\end{abstract}
\label{PACS number:}

\newpage

\section*{Introduction} Gravitational collapse has been exhaustively
studied in the last thirty years. Basically, the complexity of the field
equations is responsible for the lack of exact solutions which could
provide us with some insight to understand the collapse phenomena. For
this reason, one of the most studied simplified models is that of a
massless scalar field coupled to gravity in a spherically symmetric
context.

In several papers \cite{cris1}, Christodoulou studied in detail the
solutions of the spherically symmetric Einstein-scalar field equations
from an analytical point of view. On the other hand, an intriguing result
came from a numerical study.

More precisely, Choptuik \cite{choptuik} exhibited the occurrence of
critical phenomena in the formation of black holes by numerically
integrating the corresponding set of partial differential equations for
several initial-space-distributions of the scalar field, parameterized say
by $p$. He showed that there is a generally critical value, $p^*$, that
separates solutions containing black holes from those which do not, and
that the mass of the black hole formed near this critical point satisfies
a power law $M_{BH} \propto |p-p^*|^{\gamma}$, where $\gamma \sim 0.37$
seems to be a universal exponent. Remarkably, the same kind of critical
behavior was found for the collapse of gravitational waves in axisymmetric
spacetimes\cite{evans}, and all this may be seen as an indication that the
observed critical phenomena are independent of the collapsing matter as
well as of the symmetries considered.

Currently, it is a fascinating task to look for analytical solutions that
exhibit exactly the above-mentioned critical behavior. Husain et
al\cite{husain} obtained a result where there is, for some cases,
formation of an apparent horizon, but the black hole mass evaluated on it
grows without bound. The same situation was found by Roberts\cite{jb}
(self-similar case), and in some other models related to the self-similar
one \cite{brady}.

In this paper we analyze both the solution of the field equations in the
case treated by Roberts and the collapse phenomena for conformally coupled
massless scalar fields.

Besides several attractive features of models with conformally coupled
fields\cite{conformal}, Choptuik states that these models have a critical
behavior such as that described above\footnote{Indeed, these phenomena
occur for general non-minimally coupled scalar fields. The exponent
$\gamma$ depends weakly on the coupling constant $\xi$ of the
$\frac{1}{2}\,\xi\,\psi^2\,R$ term. The conformal case is characterized by
$\xi=1/6$.}, and our aim is to study these phenomena, but from an
analytical point of view. Our analysis is based on the technique of
generation of solutions for conformally coupled scalar fields, starting
from those associated to the ordinary scalar fields, first developed by
Bekenstein\cite{bek}.

The paper is organized as follows: in section 1 we present the basic
equations for the Einstein-scalar field system and the general solution
for the case of a simplified spherically symmetric background. It is shown
that among two possible solutions, that leading to the scalar field
collapse is the same as the one already found by Roberts. Section 2 is
devoted to the mapping of Roberts' solution into a solution for the
self-similar collapse of a conformally coupled scalar field, followed by
an analysis of the critical phenomena related to the formation of black
holes. Finally, the Conclusions contain a brief discussion about this work
and some of its possible extensions.

\section{Field Equations}

Let us consider the line element for spherically symmetric spacetimes
written as:

\begin{equation}
ds^2=h(u,v)\,du\,dv-r^2(u,v)\,(d\theta^2+sin^2\,\theta\,d\varphi^2)
\end{equation}

\noindent where $u$ and $v$ are null coordinates. The field equations in
which the ordinary massless scalar field is taken as the source of
curvature are

\begin{equation}
R_{\mu\nu}=-\phi,_{\mu}\,\phi,_{\nu}
\end{equation}

\noindent where $\phi$ is the scalar field. The Einstein-scalar field
equations are:

\begin{eqnarray}
\frac{2}{r}\,r,_{uu}-\frac{2}{r\,h}\,r,_u\,h,_u
& = & -(\phi,_u)^2
\label{fi1}
\\
\frac{2}{r}\,r,_{vv}-\frac{2}{r\,h}\,r,_v\,h,_v
& = & -(\phi,_v)^2
\label{fi2}
\\
\frac{h,_{uv}}{h}-\frac{h,_u\,h,_v}{h^2}+\frac{2}{r}\,r,_{uv}
& = & -\phi,_u\,\phi,_v
\label{fi3}
\\
(r^2),_{uv}
& = &-\frac{h}{2}
\label{r1}
\end{eqnarray}

\noindent We shall restrict our study to the case in which $h=1$. It is
not difficult to see that a general solution for eq.(\ref{r1}) can be
written as:

\begin{equation}
r^2(u,v)=-\frac{u\,v}{2}+g_1(u)+g_2(v)
\label{h1}
\end{equation}

\noindent where $g_1(u)$ and $g_2(v)$ are arbitrary functions. Now, it is
possible to explicitly determine $g_1(u)$ and $g_2(v)$ by building an
equation involving only $r$ from \eqs(\ref{fi1},\ref{fi2},\ref{fi3}),

\begin{equation}
{r,_{{vu}}}^{2}-r,_{{uu}}r,_{{vv}}=0,
\label{r2}
\end{equation}

\noindent introducing \eq(\ref{h1}) into \eq(\ref{r2}) and expanding
$g_1(u)$ and $g_2(v)$ in power series. We find, by induction, that the
latter equation restricts the form of $g_1$ and $g_2$ to second order
polynomials,

\begin{equation}
g_1(u)=a_0+a_1\,u+a_2\,u^2,\ \ \ \
g_2(v)=b_0+b_1\,v+b_2\,v^2
\label{g12}
\end{equation}
where the $a$ and $b$ coefficients must satisfy:
\begin{equation}
4\,b_{{2}}\,{a_{{1}}}^{2}
-16\,b_{{2}}\,b_{{0}}\,a_{{2}}
-16\,b_{{2}}\,a_{{0}}\,a_{{2}}
+4\,{b_{{1}}}^{2}\,a_{{2}}
+2\,b_1\,a_1
+b_{{0}}
+a_{{0}}
=0
\label{sol0}
\end{equation}

\noindent The relation above\footnote{We take $b_0=0$ without loss of
generality.} has two different solutions for $\{a_0,a_1,a_2,b_1,b_2\}$.
The first one is given by

\begin{equation}
a_0=2\,\frac{2\,a_2\,b_1^2+2\,a_1^2\,b_2+a_1\,b_1}{16\,a_2\,b_2-1}.
\label{sol1}
\end{equation}

\noindent in which case we are assuming $1-16\,a_2\,b_2 \neq 0$. Note
that, although \eq(\ref{sol1}) apparently leads to four independent
constants, $\{a_1,a_2,b_1,b_2\}$, since $1-16\,a_2\,b_2 \neq 0$ it is
always possible to remove the linear (in $u$ and $v$) and independent
terms of \eqs(\ref{h1},\ref{g12}) by introducing $u \rightarrow u-u_0$ and
$v \rightarrow v-v_0$, with suitable constant values for $u_0$ and $v_0$.
After making that shift, we arrive at a solution for $r^2(u,v)$ with only
two independent constants, which coincides with the one already found by
Roberts\cite{roberts} (self-similar case).

Another solution to \eq(\ref{sol0}) is given by:

\begin{equation}
b_{{2}}={\frac {1}{16\,a_{{2}}}},\ \ \
b_{{1}}=-{\frac {a_{{1}}}{4\,a_{{2}}}}
\end{equation}

\noindent that is,

\begin{equation}
1-16\,a_2\,b_2=0
\label{sol2}
\end{equation}

\noindent and there are only three independent constants:
$\{a_0,a_1,a_2\}$\footnote{Note that, in this case, since $1-16\,a_2\,b_2
= 0$, it is not possible to recover Roberts' solution by a constant shift
on the $u$ and $v$ variables of \eqs(\ref{g12},\ref{h1}).}.

This solution has a killing vector $k_\mu=(1/4\,b_2,1,0,0)$ besides those
associated to the spherical symmetry. In turn, this killing vector can be
timelike or spacelike, depending on the sign of $b_2$, leading to a static
or time-dependent (cosmological model) solution that cannot describe the
scalar field collapse.

Let us now solve the field equations \eqs(\ref{fi1},\ref{fi2},\ref{fi3})
for the scalar field $\phi$. Taking \eq(\ref{fi1}), for instance, along
with the solution found for $r(u,v)$ in which $1-16\,a_2\,b_2 \neq 0$, we
arrive at:

\begin{equation}
\phi(u,v)=\pm\,\frac{1}{\sqrt{2}}\,ln\,\left|
\frac{4\,a_2\,u+2\,a_1-(1+\sqrt{1-16\,a_2\,b_2})\,v
+2\,\frac{a_1+4\,a_2\,b_1}{\sqrt{1-16\,a_2\,b_2}}}{4\,a_2\,u+
2\,a_1-(1-\sqrt{1-16\,a_2\,b_2})\,v
+2\,\frac{a_1+4\,a_2\,b_1}{\sqrt{1-16\,a_2\,b_2}}}\right|,
\end{equation}

\noindent For the case in which $1-16\,a_2\,b_2 = 0$, we have:

\begin{equation}
\phi(u,v)=\pm\,\frac{1}{\sqrt{2}}\,ln\,\left|
\frac{u-2\,b_1-4\,b_2\,v-2\,\sqrt{b_1^2-4\,a_0\,b_2}}
{u-2\,b_1-4\,b_2\,v+2\,\sqrt{b_1^2-4\,a_0\,b_2}}\right|,
\end{equation}

The first case has already been analyzed\cite{jb} in order to reproduce
analytically the critical behavior obtained by Choptuik\cite{choptuik}. By
adjusting a characteristic parameter of the solution, say
$\alpha$\cite{jb} ($b_2$ in our notation), three classes are obtained:
subcritical ($0<\alpha<1/4$), where the scalar field collapses and
disperses, leaving behind a flat spacetime; critical ($\alpha=0$), where
the final result is an asymptotically flat spacetime with a null
singularity; and supercritical ($\alpha<0$), corresponding to the
formation of black holes. Among these classes, the last is the most
interesting, despite the undesirable fact that the mass of the black hole
grows without bound as $v \rightarrow \infty$.

\section{Self-Similar Solutions with Conformal Scalar Field}

In this section, we apply the technique of generating solutions for
conformally coupled scalar fields first developed by Bekenstein\cite{bek},
departing from Roberts' solution. The actions for a massless conformally
coupled scalar field $\psi$ and for the ordinary one, $\phi$, are given
respectively by

\begin{equation}
S_\psi=\int\,d^{4}\,x\,\sqrt{-g}\,(\frac{R}{2}-\frac{1}{2}\,g^{\alpha\beta}\,
\partial_{\alpha}\psi\,\partial_{\beta}\,\psi-\frac{\xi}{2}\,\psi^2\,R)
\end{equation}

\begin{equation}
S_\phi=\int\,d^{4}\,x\,\sqrt{-\tilde{g}}\,(\frac{\tilde{R}}{2}
-\frac{1}{2}\,\tilde{g}^{\alpha\beta}\,
\partial_{\alpha}\phi\,\partial_{\beta}\,\phi) \end{equation}

\noindent where $\xi=1/6$, and $\tilde{g}_{\mu\nu}$ and $g_{\mu\nu}$ are
the metric tensors associated to ordinary and conformally coupled scalar
fields, respectively. These actions are connected by the conformal
transformation:

\begin{equation}
g_{\mu\nu}=\Omega^{-2}\,\tilde{g}_{\mu\nu}=\left\{ \begin{array}{ll}
cosh^2\,(\sqrt{\xi}\,\phi)\,\tilde{g}_{\mu\nu}  \\ \\
sinh^2\,(\sqrt{\xi}\,\phi)\,\tilde{g}_{\mu\nu}
\end{array}
\right. \
\label{tr1}
\end{equation}

\noindent where $\Omega^2=|1-\xi\,\psi^2|$ and $\xi=1/6$, and the
relation between  $\psi$ and $\phi$ is given by:

\begin{equation}
\psi=\left\{ \begin{array}{ll}
\pm\,\frac{1}{\sqrt{\xi}}\,tanh\,(\sqrt{\xi}\,\phi)  \\ \\
\pm\,\frac{1}{\sqrt{\xi}}\,cotanh\,(\sqrt{\xi}\,\phi)
\end{array}
\right. \
\label{tr2}
\end{equation}

\noindent As we can see from the above, in this manner we generate two
distinct types of conformally coupled scalar field solutions. Henceforth,
we will denote the solutions for the first case type $A$ and type $B$ for
the second. Before going on with the determination and analysis of the
conformal solutions, it is useful to define the mass function $m(u,v)$ as:

\begin{equation}
m = \frac{\Sigma}{2}\,
\left( 1 + g^{\alpha\beta}\,\Sigma_{,\alpha}\,\Sigma_{,\beta} \right)
\end{equation}

\noindent where $\Sigma(u,v)$ denotes the radius of the two-sphere. From
the equation above, we can write a relation between the mass function $m$
associated to the conformally coupled scalar field solutions and the
corresponding one associated to the ordinary scalar field,
$\tilde{m}(u,v)$. After direct calculation, the following expression
arises:

\begin{equation}
\frac{m}{\Sigma}=\frac{\tilde{m}}{r}
+\frac{\Omega_{\phi}^{2}}{2\,\Omega^2}\,\tilde{g}^{\mu\nu}\,
\phi_{,\mu}\,\phi_{,\nu}\,r^2+\frac{\Omega_{\phi}}{\Omega}\,r\,
\tilde{g}^{\mu\nu}\, \phi_{,\mu}\,r_{,\nu}
\label{mass}
\end{equation}

\noindent where $\Omega_{\phi}=\frac{d\,\Omega}{d\,\phi}$ and $r(u,v)$ is
the radius of the two-sphere given by the ordinary scalar field solution.

Now, applying the transformations \eqs(\ref{tr1},\ref{tr2}) to Roberts'
solution, the metric tensor components and the conformally coupled scalar
field appear, respectively, as:

\begin{eqnarray}
h
& = &
\frac{1}{4}\,(M^{\frac{1}{2\,\sqrt{3}}} \pm M^
{\frac{-1}{2\,\sqrt{3}}})^2
\\
\Sigma^2
& = &
\frac{1}{4}\,(M^{\frac{1}{2\,\sqrt{3}}} \pm M^
{\frac{-1}{2\,\sqrt{3}}})^2\,r^2
\\
\psi
& = &
\pm\,\sqrt{6}\,\frac{1 \mp M^{\frac{-1}{\sqrt{3}}}}{1 \pm
M^{\frac{-1}{\sqrt{3}}}}
\end{eqnarray}

\noindent where $r^2=-\frac{1}{2}\,u\,v+a_2\,u^2+b_2\,v^2$ and $M$ is
given by:

\begin{equation}
M=\frac{4\,a_2\,u-(1+\sqrt{1-16\,a_2\,b_2})\,v}
{4\,a_2\,u-(1-\sqrt{1-16\,a_2\,b_2})\,v}.
\end{equation}

\noindent The solution above is self-similar since it brings about a
homothetic killing vector

$$
k_{\mu}=\frac{\Omega^{-2}}{2}\,(v,u,0,0)
$$

\noindent From these expressions and \eq(\ref{mass}), the mass function is
determined directly after some calculation as:

\begin{eqnarray}
\lefteqn{
m=}
& & \nonumber \\
& &
\frac{1}{2\,r}\,\left[
-\frac{1}{4}\,(1-16\,a_2\,b_2)\,u\,v\,-\frac{1}{12}\,(1-16\,a_2\,b_2)
\,u\,v\,
\left(\frac{M^{\frac{1}{2\,\sqrt{3}}}
\mp M^{\frac{-1}{2\,\sqrt{3}}}}{M^{\frac{1}{2\,\sqrt{3}}}
\pm
M^{\frac{-1}{2\,\sqrt{3}}}}\right)^2
\right.
\nonumber \\
& &
\left.
+\frac{1}{\sqrt{3}}\,\left(\frac{M^{\frac{1}{2\,\sqrt{3}}}
\mp
M^{\frac{-1}{2\,\sqrt{3}}}}{M^{\frac{1}{2\,\sqrt{3}}}
\pm M^{\frac{-1}{2\,\sqrt{3}}}}\right)\,
\sqrt{1-16\,a_2\,b_2}\,(b_2\,v^2-a_2\,u^2)
\right]
\end{eqnarray}

Type $A$ solutions are the ones with the upper sign in the above
expressions, whereas the lower sign case stands for what we have called
type $B$ solutions. We now choose $a_2=1/4$, without loss of generality,
in order to restrict ourselves only to the collapse case, and analyze the
following cases: (a) $0<b_2<1/4$, (b) $b_2=0$ and (c) $b_2<0$.

For type $A$ solutions, the area of the two-sphere, $4\,\pi\,\Sigma^2$,
vanishes in the same regions in which $r^2$ does, since the conformal
factor $cosh^2(\sqrt{\xi}\,\phi)$ is always different from zero. The
invariants $R^{\alpha\beta}R_{\alpha\beta}$ and
$\psi_{,\alpha}\,\psi^{,\alpha}$ diverge in these regions, indicating the
existence of singularities. Due to the fact that an everywhere conformal
transformation does not alter the spacetime structure, these solutions
exhibit the same phases as Roberts' solution.

Concerning the apparent horizon dynamics, we start from the definition of
the locus of such structures:

\begin{equation}
g^{\alpha\beta}\,\Sigma_{,\alpha}\,\Sigma_{,\beta}=0.
\end{equation}

\noindent Two relations are obtained: $\Sigma,_u=0$ and $\Sigma,_v=0$. In
neither case is it possible to express $u_{AH}$ as an explicit function of
$v$.  A numerical plot is thus necessary, indicating that the relevant
expression is $\Sigma,_v=0$, given by:

\begin{equation}
\sqrt{1-4\,b_2}\,(1-M^{\frac{-1}{\sqrt{3}}})\,v
-\sqrt{3}\,(1+M^{\frac{-1}{\sqrt{3}}})\,(-u+4\,b_2\,v)=0
\end{equation}

\noindent For $0<b_2<1/4$ and $b_2=0$, there is no apparent horizon. Only
for $b_2<0$ is the apparent horizon present (see fig. 2). It is also
possible to show that $u_{AH} \cong 2.64\,b_2\,v$, whereas for Roberts'
solution $u_{AH}=4\,b_2\,v$. Therefore, the only modification is the slope
of the curve. The mass function evaluated on the apparent horizon, i.e.
the mass of the black hole, is:

\begin{equation}
m_{AH}=\frac{1}{8}\,(M_{AH}^{\frac{1}{2\,\sqrt{3}}}+M_{AH}^
{\frac{-1}{2\,\sqrt{3}}})\,\sqrt{(1.7424\,b_2-0.32)\,b_2}\,v
\end{equation}

\noindent where
$M_{AH}=\frac{2.64\,b_2-1-\sqrt{1-4\,b_2}}{2.64\,b_2-1+\sqrt{1-4\,b_2}}$.
Again, the mass of the black hole grows without bound for $v \rightarrow
\infty$. Whether or not we consider the self-similar collapse for $0 \leq
v \leq v_0$, as done in \cite{jb}, the important thing is to exhibit the
power law of $m_{AH}$ for near-critical evolution ($b_2=0$). We obtain:

\begin{equation}
m_{AH} \cong Const.b_2^{0.21}\,v.
\end{equation}

\noindent The exponent differs from $0.37$, claimed by Choptuik and others
to be universal. In fact, we could argue that the value of the coupling
constant $\xi$ is responsible for this difference. However, as pointed out
in the numerical work, the exponent depends only weakly on $\xi$. Thus, as
in the ordinary scalar field collapse (the exponent is 0.5), the origin of
this difference is unclear within our model.

Contrary to the previous case, type $B$ solutions exhibit distinct
spacetime structure. The singular regions, in general characterized by
$\Sigma^2=0$, are described by $r^2=0$, or by
$sinh^2(\sqrt{\xi}\,\phi)=0$, i.e., the conformal factor vanishes. In this
way, an additional singular region given by $v=0$ is present. Such a
region separates two different spacetimes, which we denote by $B_+$ and
$B_-$, that are characterized by $v<0$ and $v>0$, respectively. Another
new feature is that the region $J^-$ is singular ($R^{\mu\nu}\,R_{\mu\nu}$
as well as $\psi^{,\alpha}\,\psi_{,\alpha}$ diverge on $J^-$), even though
$\Sigma^2$ is finite in it. Such a region can be interpreted as a
cosmological null singularity at the null infinity past, where the scalar
field diverges. The structure of the spacetime $B_+$ depends on $b_2$, as
we can see from fig. 2. The spacetime $B_-$ is not altered by changing the
parameter $b_2$, and, as shown in fig. 3, this spacetime is limited by
three singular regions.As before, the dynamics of the apparent horizon is
described by $\Sigma_{,v}=0$, or:

\begin{equation}
\sqrt{1-4\,b_2}\,(1+M^{-\frac{1}{\sqrt{3}}})\,u
-\sqrt{3}\,(1-M^{-\frac{1}{\sqrt{3}}})\,(4\,b_2\,v-u)=0
\end{equation}

\noindent For $B_+$, we find the same behavior as in the type $A$ solution
(or Roberts' solution): there is an apparent horizon, and consequently
black hole formation, only for $b_2<0$. The mass of the formed black hole
tends to infinity as $v \rightarrow \infty$, and for near-critical
behavior the power law displays the same value as before, i.e. $0.21$. The
structure of $B_-$ is not altered by changing the parameter $b_2$ (fig.
3); therefore, there is no critical behavior related to the black hole
formation, and the timelike singularity is surrounded by an apparent
horizon. The evolution of the mass of the ``black hole", $m_{AH}$, is
distinct from the previous case: in the very beginning ($u\,v=-\infty$),
$m_{AH}$ is infinity; then, it diminishes gradually to become equal to
zero at $v=0$.

\section{Conclusions}

We have analyzed, for the first time, self-similar collapse of conformally
coupled scalar fields. The main motivation was to reproduce analytically
the critical behavior discovered numerically by Choptuik. For this task,
we have used the technique which permits the generation of solutions for
conformally coupled scalar fields departing from those for ordinary scalar
fields. Thus, from Roberts' solution (self-similar), we obtained two types
of solutions which we denoted type $A$ and $B$.

Type $A$ solutions have no new qualitative features if compared with
Roberts' solution. The parameter $b_2$ plays the central role: the value
$b_2=0$ separates the solutions which do not form black holes
($0<b_2<1/4$) from those that do ($b_2<0$). However, as a characteristic
of the continuous self-similar regime, the mass of the black hole tends to
infinity for $v \rightarrow \infty$. Despite this undesirable and
unphysical behavior (asymptotically all spacetime becomes trapped), we
found a power law for the mass of the black hole for near critical
evolution. The exponent is $0.21$. This value is not close to $0.37$,
obtained in the numerical work, and it is not clear we can expect such a
strong influence of the coupling parameter $\xi=1/6$ on the exponent.
According to Choptuik, the exponent depends only weakly on $\xi$. It would
be interesting to check this conjecture in the more general case of a
non-minimally coupled scalar field.

Type $B$ solutions, on the other hand, reveal some new characteristics.
Due to an additional singular region described by $v=0$, two distinct
spacetimes have to be taken into account. We called them $B_+$ and $B_-$,
and they are characterized by $v>0$ and $v<0$, respectively (see figs. (2)
and (3)). The critical behavior was found to take place only for solutions
$B_+$, and the exponent for the power law associated with $m_{AH}$ is the
same as that for type $A$ solutions. As a final remark, the dynamics of
type $B_-$ solutions is independent of $b_2$: the spacetime is limited by
three singular regions (fig. (3)), and an apparent horizon encloses the
timelike singularity. Thus, this situation is not relevant with respect to
the critical behavior in the gravitational collapse.

\section{Acknowledgments} The authors acknowledge the financial support of
the Brazilian agency CNPq (Conselho Nacional de Desenvolvimento
Cient\'{\i}fico e Tecnol\'{o}gico).

\begin{itemize}

\item fig. 1 Apparent horizon for Roberts' solution, conformally coupled
scalar field collapse $A$ and $B$ for the case $b_2 < 0$.

\item fig. 2 Causal diagrams for type $B_+$ solution in the cases (a) $0 <
b_2 < 1/4$, (b) $b_2 = 0$ and (c) $b_2 < 0$. Although the region $J^-(u
\rightarrow -\infty)$ has finite area
$\Sigma^2(-\infty,v)=\frac{1}{12}\,(1-4\,b_2)\,v^2$, it is singular.

\item fig. 3 Causal diagram for $B_-$ solution. The timelike singularity
is enclosed by an apparent horizon. According with some numerical work
with respect to eq. (27) there is a region beyond the apparent horizon
where the mass function becomes negative. Then, this model must be
considered unphysical.

\end{itemize}
\end{document}